\documentclass[letterpaper]{article} 
\usepackage{aaai25}  
\usepackage{times}  
\usepackage{helvet}  
\usepackage{courier}  
\usepackage[hyphens]{url}  
\usepackage{graphicx} 
\usepackage{natbib}  
\usepackage{caption} 
\usepackage{algorithm}
\usepackage{algorithmic}

\usepackage{cite}
\usepackage{amsmath,amssymb,amsfonts}
\usepackage{algorithmic}
\usepackage{graphicx}
\usepackage{textcomp}
\usepackage{xcolor}
\usepackage[figuresright]{rotating}
\usepackage{graphicx}
\usepackage{float}  
\usepackage{subfigure}
\usepackage{multirow}

\usepackage{tikz}
\usepackage{mathtools}
\usepackage{verbatim}
\usepackage{booktabs}

\usepackage{nicefrac}       
\usepackage{microtype}      
\usepackage{newfloat}
\usepackage{listings}
\DeclareCaptionStyle{ruled}{labelfont=normalfont,labelsep=colon,strut=off} 
\floatstyle{ruled}
\newfloat{listing}{tb}{lst}{}
\floatname{listing}{Listing}
\title{Bridging the User-side Knowledge Gap in Knowledge-aware Recommendations with Large Language Models}
\author{
    Zheng Hu\textsuperscript{\rm 1},
    Zhe Li\textsuperscript{\rm 1},
    Ziyun Jiao\textsuperscript{\rm 1},
    Satoshi Nakagawa\textsuperscript{\rm 2},\\
    Jiawen Deng\textsuperscript{\rm 1}$^*$,
    Shimin Cai\textsuperscript{\rm 1}, 
    Tao Zhou\textsuperscript{\rm 1},
    Fuji Ren\textsuperscript{\rm 1}\textsuperscript{\rm 3}\thanks{Corresponding author}
}
\affiliations{
    \textsuperscript{\rm 1}School of Computer Science and Engineering, University of Electronic Science and Technology of China\\
    \textsuperscript{\rm 2}Graduate School of Information Science and Technology, University of Tokyo\\
    \textsuperscript{\rm 3}Shenzhen Institute for Advanced Study, University of Electronic Science and Technology of China\\
    \{1998huzheng, dengjw2016, ren2fuji\}@gmail.com\\
}

\usepackage{bibentry}

\begin{document}

\maketitle

\begin{abstract}
In recent years, knowledge graphs have been integrated into recommender systems as item-side auxiliary information, enhancing recommendation accuracy. However, constructing and integrating structural user-side knowledge remains a significant challenge due to the improper granularity and inherent scarcity of user-side features. Recent advancements in Large Language Models (LLMs) offer the potential to bridge this gap by leveraging their human behavior understanding and extensive real-world knowledge. Nevertheless, integrating LLM-generated information into recommender systems presents challenges, including the risk of noisy information and the need for additional knowledge transfer. In this paper, we propose an LLM-based user-side knowledge inference method alongside a carefully designed recommendation framework to address these challenges. Our approach employs LLMs to infer user interests based on historical behaviors, integrating this user-side information with item-side and collaborative data to construct a hybrid structure: the \textbf{C}ollaborative \textbf{I}nterest \textbf{K}nowledge \textbf{G}raph (CIKG). Furthermore, we propose a CIKG-based recommendation framework that includes a user interest reconstruction module and a cross-domain contrastive learning module to mitigate potential noise and facilitate knowledge transfer. We conduct extensive experiments on three real-world datasets to validate the effectiveness of our method. Our approach achieves state-of-the-art performance compared to competitive baselines, particularly for users with sparse interactions.
\end{abstract}

\begin{links}
\link{Code}{https://github.com/laowangzi/CIKGRec}
\end{links}

\section{Introduction}
\begin{figure*}[htbp]   
\begin{center}        \includegraphics[width=0.95\linewidth]{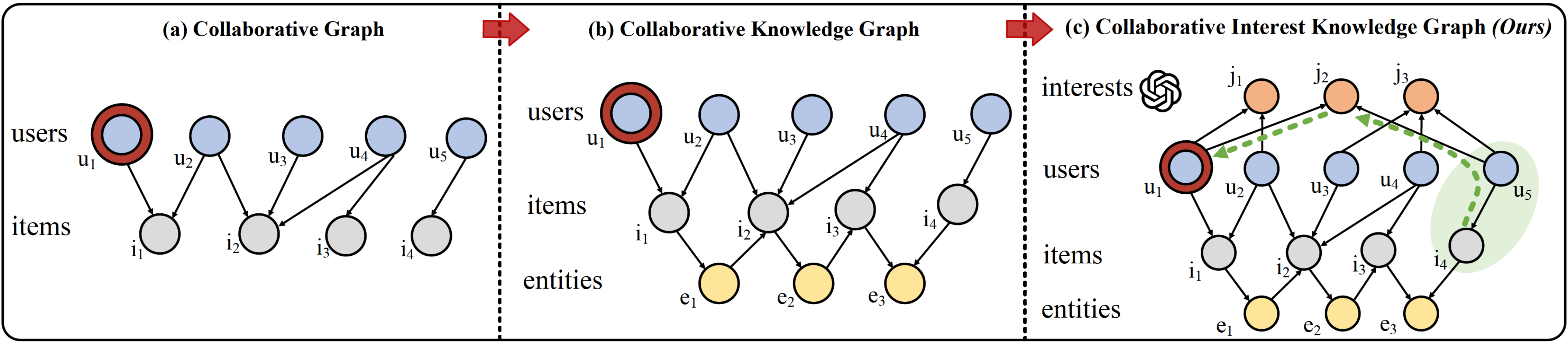}
	\caption{A toy example of the progress from traditional methods to the collaborative interest knowledge graph. $u_1$ is the target user to provide recommendation for. The green circle denote the important users and items discovered by user-side higher-order knowledge but is overlooked by traditional methods.}
	\label{fig:illustrate}
 \end{center}
\end{figure*}

In the era of information explosion, recommender systems play a crucial role in helping individuals obtain relevant information \cite{DBLP:journals/tors/GaoZLLQPQCJHL23,DBLP:conf/kdd/ChengZXTZ024}. However, these systems have long struggled with data sparsity and the cold start problem \cite{zhou2007bipartite}. Recently, knowledge graphs (KGs) have been introduced as auxiliary information on the item side, enhancing recommendation algorithms' performance and alleviating the item cold start issue \cite{DBLP:conf/recsys/CaoYWLPYY23}. Technically, KGs are integrated into the user-item collaborative graph (CG), as illustrated in Fig. \ref{fig:illustrate}(a), to form a hybrid structure known as the collaborative knowledge graph (CKG) \cite{DBLP:conf/kdd/Wang00LC19}, depicted in Fig. \ref{fig:illustrate}(b). In recent years, significant progress has been made using graph neural networks (GNNs) with the CKG for recommendations \cite{DBLP:conf/aaai/WangWX00C19,DBLP:conf/sigir/YangHXL22, DBLP:conf/kdd/YangHXH23}. However, KGs provide auxiliary information solely on the item side \cite{DBLP:conf/www/WangHWYL0C21}, and the construction and integration of auxiliary information on the user side into recommendation algorithms remain unexplored.

Unlike item-side features, which can be naturally abstracted as structured knowledge (e.g., (movie $A$, has genre, comedy)), user-side features present significant challenges in structuring and integrating into graph-based recommendation algorithms. These challenges arise from issues related to the improper granularity and scarcity of existing user-side features. Specifically, user meta-features, such as gender, age, and nationality, are too coarse to provide precise knowledge, often leading to over-smoothing problems in GNNs. While user interests are crucial for recommendations, they are typically abstract and lack explicit user feedback \cite{DBLP:journals/tkde/WuHWZW23}, making them difficult to capture and utilize.

Fortunately, the rapid development of large language models (LLMs) has shown \textit{impressive capabilities in understanding and simulating human behavior} \cite{tan2023user,DBLP:conf/cikm/HuFLHL23, DBLP:conf/www/GaoXCW0L24}. Leveraging LLMs to interpret users' historical behavior can abstract specific and meaningful user interest knowledge, thereby advancing knowledge-based recommendation algorithms. However, existing LLM-based works primarily focus on integrating recommendation tasks with dialogue systems \cite{DBLP:conf/recsys/Geng0FGZ22, DBLP:conf/ecir/HouZLLXMZ24} or using natural language features generated by LLMs as enhanced semantic information \cite{DBLP:journals/corr/abs-2402-13750,DBLP:conf/wsdm/WeiRTWSCWYH24}. How to prompt LLMs to generate and construct user-side structural knowledge and combine it with knowledge-based recommendation algorithms remains an open problem.

Further incorporating LLM-generated information into recommendation algorithms introduces \textit{two significant challenges}: (1) How to effectively utilize LLM-generated user-side knowledge while minimizing noise interference. The LLM-generated content may contain potential noise due to the hallucination problem \cite{DBLP:journals/corr/abs-2309-01219}. (2) How to effectively transfer information from the auxiliary knowledge domain to the recommendation domain. Addressing these challenges is essential for developing effective downstream recommendation algorithms.

In this paper, we propose an LLM-based user-side knowledge inference method, which generates and structures user interest information into a structural knowledge format. Additionally, we introduce a recommendation framework called \textbf{CIKGRec} that integrates the generated user-side structural knowledge into a GNN-based recommendation algorithm. Specifically, we employ LLMs to infer user interest information based on users' historical behaviors. We construct a hybrid structure, depicted in the right of Fig.~\ref{fig:illustrate}(c), called CIKG to facilitate recommendation methods in capturing user-side information. Moreover, we propose a GNN-based recommendation algorithm to effectively utilize the CIKG information. To address the issue of additional noise introduced by LLMs, we develop an enhanced graph masked autoencoder (GMAE)-based user interest reconstruction module to improve model robustness. To address the challenge of knowledge transfer across domains, we develop a cross-domain contrastive learning module that maximizes mutual information between the auxiliary information domain and the recommendation domain. Extensive experiments on three real-world datasets demonstrate that our method outperforms baseline methods, particularly for users with sparse interactions. The main contributions of this paper are as follows:
\begin{itemize}
  \item We propose an LLM-based user-side knowledge construction method, which enhances the existing CKG by integrating user interest knowledge, resulting in a novel hybrid structure called CIKG. This enrichment provides additional auxiliary information, which is beneficial for recommendations from a graph structure perspective.
  \item We further introduce a CIKG-based recommendation framework that effectively addresses the noise and information transfer challenges posed by incorporating LLM-generated user-side information into the recommendation domain.
  \item To verify the effectiveness of our proposed method, we conduct extensive experiments on three real-world datasets. Our method achieves optimal results overall and is particularly effective in alleviating the data sparsity problem for users.
\end{itemize}

\begin{figure*}[htbp]   
\begin{center}      
\includegraphics[width=1.0\textwidth]{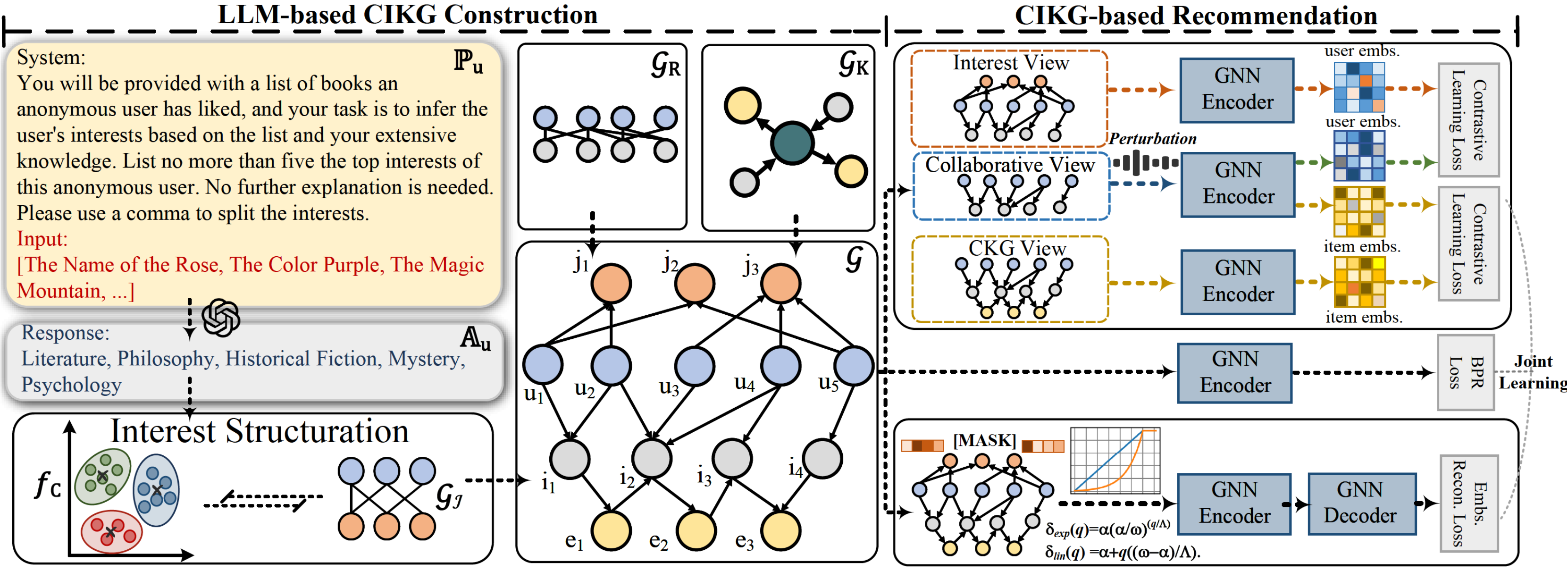}
	\caption{The overall architecture of our method. The left half illustrates the LLM-based user-side knowledge construction, and the right half shows the CIKG-based recommendation.}
	\label{fig:framework}
 \end{center}
\end{figure*}

\section{Related Work}

\textbf{LLM for Recommender Systems.} In recent years, the rapid development of LLMs has garnered increasing attention for their application in recommendation tasks. The role of LLMs in these tasks can be categorized into two primary functions \cite{DBLP:journals/corr/abs-2307-02046}: LLMs as recommenders and LLMs as enhancers. When used as recommenders, LLMs integrate recommendation tasks into conversational scenarios, tokenizing item IDs and user IDs. Due to their extensive real-world knowledge, LLMs as recommender exhibit advantages in few-shot scenarios \cite{DBLP:conf/sigir/LiR0A0S24}, fairness \cite{DBLP:conf/www/JiangBZW0F024} and explainability \cite{DBLP:journals/corr/abs-2303-14524, xie2023factual}. However, they struggle to capture the pervasive collaborative signals, resulting in limited advantages over traditional collaborative filtering algorithms \cite{DBLP:journals/corr/abs-2305-06474}. Some studies have explored enhancing traditional recommendation algorithms by incorporating features generated by LLMs. These models generate missing item features \cite{DBLP:conf/sigir/RenCYLJCZM024}, user-item interactions \cite{DBLP:conf/wsdm/WeiRTWSCWYH24}, etc., and encode LLM-generated context from natural language into semantic vectors for enhancing traditional algorithms. Nevertheless, current methods of using LLMs for data enhancement primarily focus on the meta-features, neglecting structural knowledge from the user side. This limitation hinders their ability to facilitate traditional methods to capture higher-order collaborative information.

\noindent\textbf{KG-based Recommendation Methods.} Another closely related area of this work involves KG-based recommendation methods. KB4Rec \cite{DBLP:journals/dint/ZhaoHYDHOW19} links knowledge base information with items in the recommendation domain, constructing an item-side KG to facilitate the fusion of KG and the recommendation task. Existing KG-based methods can be broadly categorized into two groups: path-based methods and regularization-based methods. Path-based approaches extract higher-order information paths and input them into recommendation models \cite{DBLP:conf/aaai/WangWX00C19, DBLP:conf/kdd/HuSZY18}. Regularization-based techniques introduce additional loss terms to capture the KG's structure, regulating the learning process of recommender models \cite{DBLP:conf/kdd/ZhangYLXM16, DBLP:conf/kdd/Wang00LC19,DBLP:conf/www/WangZXLG19,DBLP:conf/www/WangHWYL0C21,DBLP:journals/asc/HuCWZ23}. Recent efforts have focused on enhancing model robustness \cite{DBLP:conf/sigir/YangHXL22} and reconstructing knowledge graphs \cite{DBLP:conf/kdd/YangHXH23, DBLP:conf/sigir/TangGW0WFZ24} with information from the recommendation domain. However, existing methods often overlook structured knowledge on the user side and fail to process unreliable content generated by LLMs.

\section{Problem Definition}

In this section, we first introduce the concept of the CG and the KG. Then, we formally define the top-$K$ recommendation task.

\noindent\textbf{Definition of CG.} In a recommendation scenario, historical user-item interactions (e.g., watches and clicks), denoted as $\mathcal{E}^+$, are represented as a user-item collaborative graph $\mathcal{G}_R=(\mathcal{V}_R, \mathcal{E}^+)$, where $\mathcal{V}_R=(\mathcal{U} \cup \mathcal{I})$, with $\mathcal{U}$ representing the set of users and $\mathcal{I}$ representing the set of items.

\noindent\textbf{Definition of KG.} In addition to interactions, item-side information is organized as a knowledge graph $\mathcal{G}_K$, a directed graph composed of subject-property-object triples \cite{DBLP:conf/aaai/ShiW17}. Formally, $\mathcal{G}_K = \{(h, r, t) | h, t \in \mathcal{V}_E, r \in \mathcal{R}\}$, where $\mathcal{V}_E$ denotes the set of entities and $\mathcal{R}$ denotes the set of relations. An item-entity projection set $\mathcal{P}=\{(i, e) | i \in \mathcal{I}, e \in \mathcal{V}_E\}$ links items to entities.

\noindent\textbf{Top-$K$ Recommendation.}
The top-$K$ recommendation task is defined as follows: Given $\mathcal{G}_R$, $\mathcal{G}_K$ and $\mathcal{E}^+$, the goal is to recommend the top-$K$ items that each user $u \in \mathcal{U}$ is most likely to interact with, among the items $u$ has not yet interacted with.

\section{Method}
The architecture of our approach is depicted in Fig.~\ref{fig:framework}. Overall, our method comprises two stages: LLM-based CIKG construction (left half of Fig.~\ref{fig:framework}) and CIKG-based recommendation (right half of Fig.~\ref{fig:framework}). The details of these stages are as follows.

\subsection{LLM-based CIKG Construction}

This section introduces the details of LLM-based user-side knowledge generation and how to convert such knowledge from natural language into structured user-side information. Furthermore, we introduce the concept of the CIKG.

\subsubsection{LLM-based User Interest Inference.} Modeling and representing user interests is crucial for recommender systems \cite{DBLP:conf/aaai/GaoTHL15,DBLP:conf/www/YangHXHLL23}. Traditional recommendation algorithms represent user interests implicitly as dense vectors \cite{DBLP:journals/tkde/WuHWZW23}. The abstract and complex nature of user interests makes explicitly representing them challenging. This limitation hinders the recommendation model's ability to fully leverage user interest information. Recently, LLMs have shown good ability to understand user behavior \cite{tan2023user,DBLP:conf/cikm/HuFLHL23, DBLP:conf/www/GaoXCW0L24}, which makes it possible to model user interests explicitly. Inspired by this, we devise prompts derived from the dataset's user behaviors to enable the LLM to infer and generate user interests not originally present in the dataset. Leveraging the LLM's extensive knowledge and reasoning ability, the model can summarize high-level, abstract user interests based on the relevance of items interacted with by users rather than relying on traditional user-side meta-features. Specific examples are shown in Fig.~\ref{fig:framework}. Formally, the LLM-based user interest inference is defined as follows:
\begin{equation}
  \mathbb{A}=\{\mathbb{A}_1, \mathbb{A}_2, ..., \mathbb{A}_{|\mathcal{U}|}\},\ \mathbb{A}_u = LLM(\mathbb{P}_u, \mathcal{E}^+_u),
\end{equation}
where $\mathbb{P}_u$ represents the textual prompt for user $u$, and $\mathbb{A}$ represents the generated user interest textual information.

\subsubsection{User Interest Knowledge Structuration.} The purpose of structuring user interest knowledge is twofold: (1) Ensuring vector space consistency. Previous works \cite{DBLP:conf/sigir/RenCYLJCZM024,DBLP:conf/wsdm/WeiRTWSCWYH24} have encoded LLM-generated text as semantic embeddings and incorporated them into traditional recommendation methods. However, direct addition poses the problem of misalignment \cite{DBLP:conf/mm/HaoWGLW19, DBLP:conf/icml/RadfordKHRGASAM21} between semantic signals and collaborative signals, resulting in suboptimal performance. (2) Capturing user-side multi-hop information. Explicitly modeling user interests as structured information facilitates the model mining higher-order relationships between users with common interests. Technically, we first employ text clustering methods, which can be as simple as tf-idf or deep sentence embedding methods, to merge similar interests. In this way, we semantically avoid inconsistencies where the same interests are expressed differently due to the randomness of LLM outputs. We then connect the merged interests with users to build a user interest knowledge graph $\mathcal{G}_J$:
\begin{equation}
\begin{aligned}
  \mathcal{J}&=\{j_1, j_2,...j_{\kappa}\}=f_C(\mathbb{A}, \kappa), \\
  \mathcal{G}_J&=(\mathcal{V}_J,\mathcal{E}_J), \ \mathcal{V}_J=(\mathcal{U}\ \cup\ \mathcal{J}),
\end{aligned}
\end{equation}
where $f_C(\cdot)$ denotes the textual clustering method, and $\kappa$ is a hyperparameter representing the number of clusters. An edge $(u, j) \in \mathcal{E}_J$ exists if user $u$ has the interest $j$.

The CIKG is defined as a unified relational graph $\mathcal{G}$, encoding user behaviors, user interest knowledge, and item-side knowledge. Each user behavior in $\mathcal{G}_R$ is represented as a triplet $(u, \textit{Interact}, i)$, where \textit{Interact} denotes the relationship between user $u$ and item $i$. User interest information in $\mathcal{G}_J$ is represented as triplets $(u, \textit{Has Interest}, j)$. By integrating the item-entity projection set $\mathcal{P}$, the user behaviors graph $\mathcal{G}_R$, and user-side information ${\mathcal{G}_J}$ with the knowledge graph $\mathcal{G}_K$, we form the unified graph $\mathcal{G}=\{(h, r, t)|h,t\in \mathcal{V}=(\mathcal{U} \cup \mathcal{V}_E \cup \mathcal{J}), r \in \mathcal{R}'\}$, where $\mathcal{R}'=\mathcal{R} \cup \{Interact, Has\ Interest\}$, as the CIKG. Let $Z \in \mathbb{R}^{|\mathcal{V}|\times D}$ denotes the initial embedding matrix of $\mathcal{G}$, where $D$ is the embedding dimension. $z_i \in \mathbb{R}^{D \times 1}$ represents the embedding of node $i$.

\subsection{CIKG-based Recommendation}
In this section, we propose a CIKG-based recommendation method. This method consists of three components: user interest reconstruction \textit{(addressing the challenge of LLM-generated noisy content)}, cross-domain contrastive learning \textit{(addressing the challenge of knowledge transfer)}, and model training. The details are as follows.

\subsubsection{User Interest Reconstruction.} Due to the hallucination problem \cite{DBLP:journals/corr/abs-2309-01219} in LLMs and the complexity of user interests, the user interests inferred by LLMs can be inaccurate and noisy. The objective of this module is to reduce the noise interference. Inspired by the robustness of graph representations achieved through GMAEs, we apply masking and reconstruction on user interest nodes, which facilitates the model's focus on crucial information and enhances its generalization ability. Unlike previous GMAE methods that use a fixed mask rate \cite{DBLP:conf/kdd/HouLCDYW022} or incorporate randomness into the mask rate determination \cite{DBLP:conf/aaai/0001T0W024}, we propose two mask rate acceleration scheduling strategies that dynamically control the mask rate. Drawing inspiration from curriculum learning \cite{DBLP:journals/pami/WangCZ22}, these strategies progressively transition task difficulty from easy to hard. Formally, the two strategies can be expressed as:
\begin{equation}
\begin{aligned}
  p^q\! =\! \psi(\delta(q),\ \omega), \ \delta(q)\! =\! \left\{
  \begin{aligned}
  &\delta_{lin}(q)\! =\! \alpha+q\frac{(\omega-\alpha)}{\Lambda},\\
  &\delta_{exp}(q)\! =\! \alpha(\frac{\omega}{\alpha})^{\frac{q}{\Lambda}},
   \end{aligned}
   \right.\\
\end{aligned}
\end{equation}
where $\delta(\cdot)$ denotes the scheduling function. $\alpha$, $\omega$, and $\Lambda$ are hyperparameters representing the initial and maximum mask rates, and the required epochs to convergence to the maximum mask rate. $q$ denotes the current epoch number. $p^q$ denotes the mask rate at epoch $q$. $\psi(\cdot)$ is the minimize function. It's easy to prove that: 
\begin{equation}
\delta_{exp}(q) < \delta_{lin}(q), 0 < q < \Lambda.
\end{equation}
Therefore, our proposed exponential growth strategy allows the task difficulty to grow more slowly, as shown in the user interest reconstruction module of Fig.~\ref{fig:framework}. After determining the mask rate $p^q$, we sample a subset of nodes $\widetilde{\mathcal{J}} \in \mathcal{J}$ and mask their embeddings with a learnable mask token $[M]$. For each interest node $j$, we have:
\begin{equation}
  z'_j=\left\{
  \begin{aligned}
  &z_{[M]}\ \ \ &if\ j \in \widetilde{\mathcal{J}},\\
  &z_j \ \ \ &if\ j \notin \widetilde{\mathcal{J}}.\\
   \end{aligned}
   \right.
\end{equation}
The masked embeddings $Z'$ and $\mathcal{G}$ are then sent to the encoder $f_{G}^{l-1}(\cdot)$, which is a GNN with $(l-1)$ layers. Without loss of generality, we use LightGCN \cite{DBLP:conf/sigir/0001DWLZ020} as the graph encoder for all modules. The decoder $f_G^1(\cdot)$ is then applied to obtain the reconstructed node embeddings. This process can be formulated as:
\begin{equation}
  \begin{aligned}
  Z''\!= f_G^{l-1}(Z',\ \mathcal{G}),\ \ \ 
  Z'''\!= f_G^{1}(\chi(Z''),\ \mathcal{G}),
   \end{aligned}
\end{equation}
where $\chi(\cdot)$ denotes the linear projection layer. Subsequently, we define the user interest reconstruction loss by comparing $Z$ and $Z'''$. The loss function, with the scaling factor $\eta$, is described as follows:
\begin{equation}
  \begin{aligned}
  \mathcal{L}_u = \frac{1}{|\widetilde{\mathcal{J}}|}\sum\limits_{j \in \widetilde{\mathcal{J}}}(1-\frac{z^T_j z'''_j}{\Vert z_j \Vert \cdot \Vert z'''_j \Vert})^\eta,\ \eta \geq 1,
   \end{aligned}
\end{equation}
where $\Vert \cdot \Vert$ represents the $L_2$ normalization function.

\begin{table*}[!t]
  \centering
  \caption{The overall performance of
 CIKGRec and baseline models. The best and second-best performances are
 highlighted in \textbf{bold} and \underline{underline}, respectively.}
  \setlength{\tabcolsep}{4pt}
  \renewcommand\arraystretch{1.0}
  \resizebox{\textwidth}{!}{
    \normalsize
    \begin{tabular}{c|cccc|cccc|cccc}
    \hline\rule{0pt}{10pt}
    \multirow{2}[1]{*}{\centering MODEL} & \multicolumn{4}{c|}{DBbook2014} & \multicolumn{4}{c|}{Book-Crossing} & \multicolumn{4}{c}{MovieLens-1M} \\
\cline{2-13}\rule{0pt}{10pt}          & R@50  & R@100 & N@50  & N@100 & R@50  & R@100 & N@50  & N@100 & R@50  & R@100 & N@50  & N@100 \\
    \hline
    \hline\rule{0pt}{10pt}
    LightGCN & 0.4214  & 0.5358  & 0.2230  & 0.2481  & 0.1607  & \underline{0.2246}  & 0.0765  & 0.0911  & 0.4086  & 0.5574  & 0.3977  & 0.4432  \\
    SGL   & 0.4258  & 0.5249  & 0.2350  & 0.2567  & 0.1613  & 0.2154  & 0.0860  & 0.0982  & 0.4094  & 0.5505  & 0.4004  & 0.4433  \\
    SimGCL & 0.4280  & 0.5369  & 0.2376  & 0.2610  & \underline{0.1653}  & 0.2211  & 0.0860  & 0.0988  & 0.4131  & 0.5560  & 0.4048  & 0.4482  \\
    \hline\rule{0pt}{10pt}
    CKE   & 0.3419  & 0.4448  & 0.1832  & 0.2060  & 0.1284  & 0.1691  & 0.0708  & 0.0811  & 0.3948  & 0.5421  & 0.3841  & 0.4295  \\
    CFKG  & 0.3586  & 0.4559  & 0.1627  & 0.1816  & 0.1216  & 0.1771  & 0.0503  & 0.0617  & 0.3930  & 0.5439  & 0.3672  & 0.4142  \\
    \hline\rule{0pt}{10pt}
    KGAT  & 0.4174  & 0.5309  & 0.2110  & 0.2360  & 0.1564  & 0.2133  & \underline{0.0865}  & \underline{0.1004}  & 0.4059  & 0.5527  & 0.3928  & 0.4378  \\
    KGIN  & 0.4147  & 0.5337  & 0.2200  & 0.2462  & 0.1468  & 0.2075  & 0.0681  & 0.0819  & 0.4095  & 0.5581  & 0.4004  & 0.4453  \\
    KGCL  & 0.4308  & 0.5271  & \underline{0.2453}  & \underline{0.2666}  & 0.1562  & 0.2049  & 0.0859  & 0.0971  & 0.4083  & 0.5496  & 0.3996  & 0.4430  \\
    MCCLK & 0.4376  & 0.5491  & 0.2054  & 0.2273  & 0.1476  & 0.2039  & 0.0656  & 0.0774  & \underline{0.4176}  & \underline{0.5655}  & 0.3876  & 0.4339  \\
    KGRec & \underline{0.4415}  & \underline{0.5497}  & 0.2373  & 0.2611  & 0.1473  & 0.2075  & 0.0699  & 0.0836  & 0.4136  & 0.5633  & \underline{0.4072}  & \underline{0.4524}  \\
    \hline\rule{0pt}{10pt}
    \textbf{CIKGRec} & \textbf{0.4642} & \textbf{0.5756} & \textbf{0.2494} & \textbf{0.2739} & \textbf{0.1791} & \textbf{0.2455} & \textbf{0.0889} & \textbf{0.1041} & \textbf{0.4250} & \textbf{0.5718} & \textbf{0.4162} & \textbf{0.4609} \\
    \hline\rule{0pt}{10pt}
    \textit{\%Imp.} & 5.14\% & 4.71\% & 1.67\% & 2.74\% & 8.35\% & 9.31\% & 2.77\% & 3.69\% & 1.77\% & 1.11\% & 2.21\% & 1.88\% \\
    \hline
    \end{tabular}%
    }
  \label{tab:overall_performance}%
\end{table*}%

\begin{table}[htbp]
  \centering
  \normalsize
  \tabcolsep=0.6mm
  \caption{Statistics of the datasets.}
  \renewcommand\arraystretch{1.05}
    \begin{tabular}{l|r|r|r}
    \hline
    Dataset & \multicolumn{1}{|c}{DBbook2014} & \multicolumn{1}{|c}{Book-Crossing} & \multicolumn{1}{|c}{MovieLens-1M} \\
    \hline
    \#Users & 5,576 & 6,616 & 6,040 \\
    \#Items & 2,680 & 8,853 & 3,260 \\
    \#Ratings & 65,961 & 110,662 & 998,539 \\
    \#Density & 0.44\%  & 0.19\%  & 5.07\% \\
    \hline 
    \#Relations & 13 & 4 & 20 \\
    \#Entities & 8,762 & 1,404 & 14,377 \\
    \#Triplets & 134,223 & 1,137 & 415,104 \\    
    \hline
    \end{tabular}%
  \label{tab:statistics_datasets}%
\end{table}%

\subsubsection{Cross-domain Contrastive Learning.} To facilitate the transfer of user-side and item-side knowledge to the recommendation domain, we devise a cross-domain contrastive learning module. Unlike most existing contrastive learning-based models \cite{DBLP:conf/nips/YouCSCWS20,DBLP:conf/sigir/WuWF0CLX21} that augment graphs by perturbing the graph structure and potentially losing useful information, we construct augmented views by adding auxiliary information from the user or item side to $\mathcal{G}_R$. Our proposed augmentation approach aligns auxiliary information with the collaborative signals of the recommendation domain, all while preserving the integrity of the graph structure. The encoded embeddings of the augmented views are as follows:
\begin{equation}
\begin{aligned}
  &Z^{v1} = f^l_G(Z,\ \mathcal{G}_{R},\ \gamma),\\
  &Z^{v2} = f^l_G(Z,\ \mathcal{G}_{v2}),\ \ \mathcal{G}_{v2}\!=\!Merge(\mathcal{G}_{R},\ \mathcal{G}_J),\\
  &Z^{v3} = f^l_G(Z,\ \mathcal{G}_{v3}),\ \ \mathcal{G}_{v3}\!=\!Merge(\mathcal{G}_{R},\ \mathcal{M}(\mathcal{G}_K,\mathcal{P})),\\
\end{aligned}
\end{equation}
where $\gamma$ is representation-level perturbations as depicted in SimGCL \cite{DBLP:conf/sigir/YuY00CN22}, which provides augmentation to $\mathcal{G}_R$ without perturb graph structures. $Merge(\cdot)$ is a function to merge the edges and nodes of graphs and eliminate co-reference nodes or edges. $\mathcal{M}(\cdot)$ is the map function to transfer entity id to item id based on $\mathcal{P}$. 

The positive samples of the user node in $\mathcal{G}_{v1}$ are the user node in $\mathcal{G}_{v2}$, and the positive examples of the item node in $\mathcal{G}_{v1}$ are the item nodes in $\mathcal{G}_{v3}$, since $\mathcal{G}_{v2}$ and $\mathcal{G}_{v3}$ add the knowledge of the user- and item-side, respectively. The contrastive objective $\mathcal{L}_c$, which formally maximizes agreement among positive pairs and minimizes agreement among negative pairs, is defined based on the InfoNCE loss \cite{DBLP:conf/icml/ChenK0H20} as follows:
\begin{equation}
\begin{aligned}
\mathcal{L}_c = &\sum\limits_{u\in\mathcal{U}}-log\frac{exp(s(z_u^{v1},z_u^{v2})/\tau)}{\sum\limits_{w\in\mathcal{U}}exp(s(z_u^{v1},z_w^{v2})/\tau)} +\\ 
    &\sum\limits_{i\in\mathcal{I}}-log\frac{exp(s(z_i^{v1},z_i^{v3})/\tau)}{\sum\limits_{w\in\mathcal{I}}exp(s(z_i^{v1},z_w^{v3})/\tau)},
\end{aligned}
\end{equation}
where $s(\cdot)$ is the cosine function to estimate the similarity of positive and negative pairs. $\tau$ is the temperature coefficient to control the softness of the probability distribution.

\subsubsection{Model Training.}
In addition to the two loss functions introduced above, the model training process also includes the recommendation task loss function $\mathcal{L}_r$, which can be formulated as:
\begin{equation}
\begin{aligned}
&\hat{Z} = f_G^l(Z, \mathcal{G}), \\
\mathcal{L}_r = \sum\limits_{(u,i,i')}&-ln(\sigma(\hat{z}^T_u\hat{z}_{i} - \hat{z}^T_u\hat{z}_{i'})),
\end{aligned}
\end{equation}
where $(u, i) \in \mathcal{E}^+$ and $i'$ is a randomly sampled negative item of user $u$. $\sigma$ is the sigmoid function. The main loss is the weighted sum of the three loss functions, which can be formulated as:
\begin{equation}
\begin{aligned}
\mathcal{L} = \lambda_1 \mathcal{L}_r + \lambda_2 \mathcal{L}_u + \lambda_3 \mathcal{L}_c,
\end{aligned}
\end{equation}
where $\lambda$ is the hyperparameter to control the measures of loss functions. In order to further enhance the multi-relational semantic representation space for entity-item dependencies, we perform the alternative translation-based training \cite{DBLP:conf/nips/BordesUGWY13}, which can be formulated as:
\begin{equation}
\begin{aligned}
\mathcal{L}_t\! =\! \sum\limits_{(h,r,t,t')}-ln\sigma(\Vert z_h + z_r - z_{t'} \Vert - \Vert z_h + z_r - z_t \Vert),
\end{aligned}
\end{equation}
where $(h, r, t) \in \mathcal{G}_K$ and the negative tail $t'$ of head $h$ is randomly sampled from $\mathcal{G}_K$.

\section{Experiments}
\subsection{Experimental Settings}
\subsubsection{Datasets.} We conduct extensive experiments on three real-world datasets: DBbook2014 \cite{DBLP:conf/www/0003W0HC19}, Book-Crossing \cite{DBLP:conf/aaai/DongYWSYZ17}, and MovieLens-1M \cite{DBLP:journals/tist/NoiaOTS16}, which all contains item-side auxiliary knowledge and vary in terms of domain, size, and density. We follow the previous literature \cite{DBLP:conf/www/0003W0HC19,10771708,li2023exploring} filtering out low-frequency users (i.e., lower than $10$ in movielens and $5$ in DBbook2014 and Book-Crossing). The statistics of the processed datasets are shown in Tab.~\ref{tab:statistics_datasets}. 

\subsubsection{Baselines.} We compare the performance of CIKGRec with various baseline methods, which can be broadly categorized into three groups: \textbf{(1) GNN-based general recommenders}, including LightGCN \cite{DBLP:conf/sigir/0001DWLZ020} and two contrastive learning-based methods, SGL \cite{DBLP:conf/sigir/WuWF0CLX21} and SimGCL \cite{DBLP:conf/sigir/YuY00CN22}. \textbf{(2) Embedding-based knowledge-aware recommenders}, including CKE \cite{DBLP:conf/kdd/ZhangYLXM16} and CFKG \cite{DBLP:journals/algorithms/AiACZ18}. \textbf{(3) GNN-based item-side knowledge-aware recommenders}, including KGAT \cite{DBLP:conf/kdd/Wang00LC19}, KGIN \cite{DBLP:conf/www/WangHWYL0C21}, and self-supervised learning-enhanced methods, KGCL \cite{DBLP:conf/sigir/YangHXL22}, MCCLK \cite{DBLP:conf/sigir/Zou0MWQ0C22}, KGRec \cite{DBLP:conf/kdd/YangHXH23}.

\subsubsection{Implementation Details and Evaluation Metrics.} We construct the training, validation, and test sets by randomly splitting the historical behavior of each user in a ratio of $7:1:2$. For our model, we search the learning rate from $0.0001$ to $0.005$. We test the user interest reconstruction initial mask rate $\alpha$ from $0.05$ to $0.15$ and the maximum mask rate $\omega$ from $0.35$ to $0.95$. In this paper, we adopt the ``gpt-3.5-turbo-0125" model as the LLM. The required convergence epoch $\Lambda$ is chosen in $\{80, 160, 320\}$. For all baseline models, we utilize the source code provided by the original authors. Hyperparameters for these baselines are either adopted as recommended by the original authors or fine-tuned using the validation sets. To ensure a fair comparison, we maintain consistent common hyperparameters across all models, including an embedding dimension of $64$ and an evaluation interval of $1$ on all datasets. An early stopping strategy with a maximum of $2000$ epochs is applied to both the baselines and our model. For all methods, we report the mean and standard deviation over $10$ runs with different random seeds. To evaluate the performance of our model on the top-$K$ recommendation task, we employ two widely used metrics: Recall@$K$ and NDCG@$K$. These metrics respectively focus on measuring the model's ability to retrieve relevant items and its ability to rank items. In this paper, $K$ takes the values of $50$ and $100$.

\begin{figure}[t]   
\begin{center}      
\includegraphics[width=0.472\textwidth]{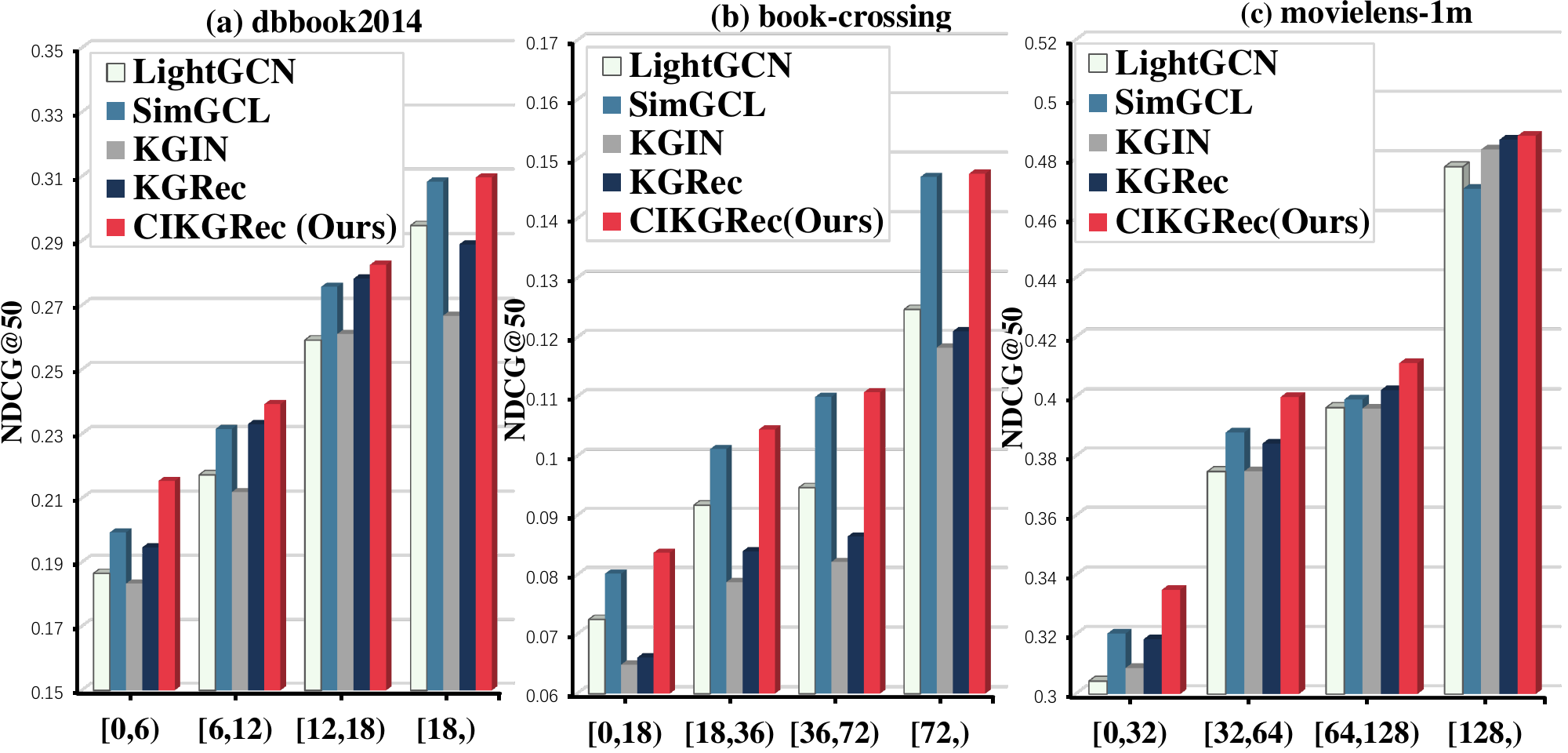}
	\caption{Performance comparisons under different sparsity user groups.}
	\label{fig:user_cold}
 \end{center}
\end{figure}

\subsection{Overall Performance}
The overall performance of CIKGRec and baseline models is shown in Tab.~\ref{tab:overall_performance}. Based on the results, we have the following key observations: \textbf{(1)} Our model achieves superior performance across all metrics and datasets, demonstrating the effectiveness of our model. This also underscores the efficacy of the proposed user interest reconstruction module and cross-domain contrastive learning module. \textbf{(2)} The model shows notable improvements on the two data-sparse datasets (DBbook2014 and Book-Crossing), with densities of $0.44\%$ and $0.19\%$, respectively. We attribute this to our proposed LLM-generated user-side structural knowledge, which effectively mitigates the data sparsity problem. \textbf{(3)} We observe that the knowledge-aware baselines are sensitive to the availability of item-side knowledge. Specifically, these methods perform well in datasets with abundant item-side knowledge (DBbook2014 and MovieLens-1M) but perform worse in datasets with sparse item-side knowledge (Book-Crossing). Our method performs well on both item-side knowledge-dense and knowledge-sparse datasets, highlighting the effectiveness and necessity of incorporating user-side knowledge.

\subsection{User-side Data Sparsity Analysis}
One motivation for exploiting user-side knowledge is to alleviate the data sparsity problem, which often limits the expressiveness of recommender systems. It is challenging to learn optimal representations for inactive users with few interactions. Following previous work \cite{DBLP:conf/kdd/Wang00LC19}, we divide users into four groups based on their historical interactions in the training set. We then report the average $NDCG@50$ results for each user group using our model and selected baseline methods from various categories. The results are depicted in Fig.~\ref{fig:user_cold}, where the horizontal axis represents user interaction from sparse to dense.

Fig.~\ref{fig:user_cold} indicates that our model demonstrates superior performance for user groups with sparse interactions across the three datasets. This highlights the advantage of our model in handling user interaction sparsity. We attribute this superiority to our proposed CIKG, which incorporates LLM-generated user-side knowledge. The introduction of user-side knowledge effectively facilitates the mining of user-side collaborative information, thereby improving performance for user groups with sparse interactions. This also demonstrates the effectiveness of our user-side knowledge construction method.

\begin{table}[t]
  \centering
  \tabcolsep=0.3mm
  \renewcommand\arraystretch{1.0}
  \caption{Ablation study of CIKGRec.}
    \begin{tabular}{lcccccc}
    \toprule
          & \multicolumn{2}{c}{DBbook2014} & \multicolumn{2}{c}{Book-Crossing} & \multicolumn{2}{c}{MovieLens-1M} \\
    \midrule
          & \multicolumn{1}{c}{R@100} & \multicolumn{1}{c}{N@100} & \multicolumn{1}{c}{R@100} & \multicolumn{1}{c}{N@100} & \multicolumn{1}{c}{R@100} & \multicolumn{1}{c}{N@100} \\
          \cmidrule(r){2-3} \cmidrule(r){4-5} \cmidrule(r){6-7}
    Ours  & \textbf{0.5756} & \textbf{0.2739} & \textbf{0.2455} & \textbf{0.1041} & \textbf{0.5718} & \textbf{0.4609} \\
    w/o UIK & 0.5673 & 0.2615 & 0.233 & 0.0991 & 0.5662 & 0.4527 \\
    w/o UIR & 0.5701 & 0.2704 & 0.2081 & 0.0843 & 0.5713 & 0.4601 \\
    w/o CL & 0.5732 & 0.2659 & 0.2291 & 0.0917 & 0.5583 & 0.4480 \\
    \midrule
    w/o DMR & 0.5712 & 0.2709 & 0.2129 & 0.0853 & 0.5710 & 0.4596 \\
    w linear & 0.5733 & 0.2716 & 0.2452 & 0.1036 & 0.5714 & 0.4603 \\
    \bottomrule
    \end{tabular}%
  \label{tab:ablation}%
\end{table}%

\subsection{Ablation Study}

To evaluate the contributions of the various modules in our model, we conducted ablation experiments while keeping the hyperparameters fixed. The results, presented in Table 3, highlight the impact of each module. 

First, we observe a significant performance decrease across all three datasets in the model without user interest knowledge (UIK). This underscores the importance of the LLM-generated user-side structured knowledge. Similarly, the removal of user interest reconstruction (UIR) and contrastive learning (CL) leads to diminished performance, indicating these modules' roles in mitigating information noise and enhancing knowledge transfer. To investigate the impact of the dynamic mask rate strategy within the UIR, we compare two variants: one without the dynamic masking rate (DMR), and one using the linear strategy. Our findings are twofold: \textbf{(1)} Both the linear and exponential growth strategies outperform the fixed masking rate, demonstrating that an easy-to-hard training approach enhances model performance. \textbf{(2)} The exponential growth strategy yields superior results compared to the linear strategy, suggesting that a slower increase in model difficulty improves performance across the datasets, validating the efficacy of our proposed exponential growth strategy.

\begin{figure}[t]   
\begin{center}      
\includegraphics[width=0.48\textwidth]{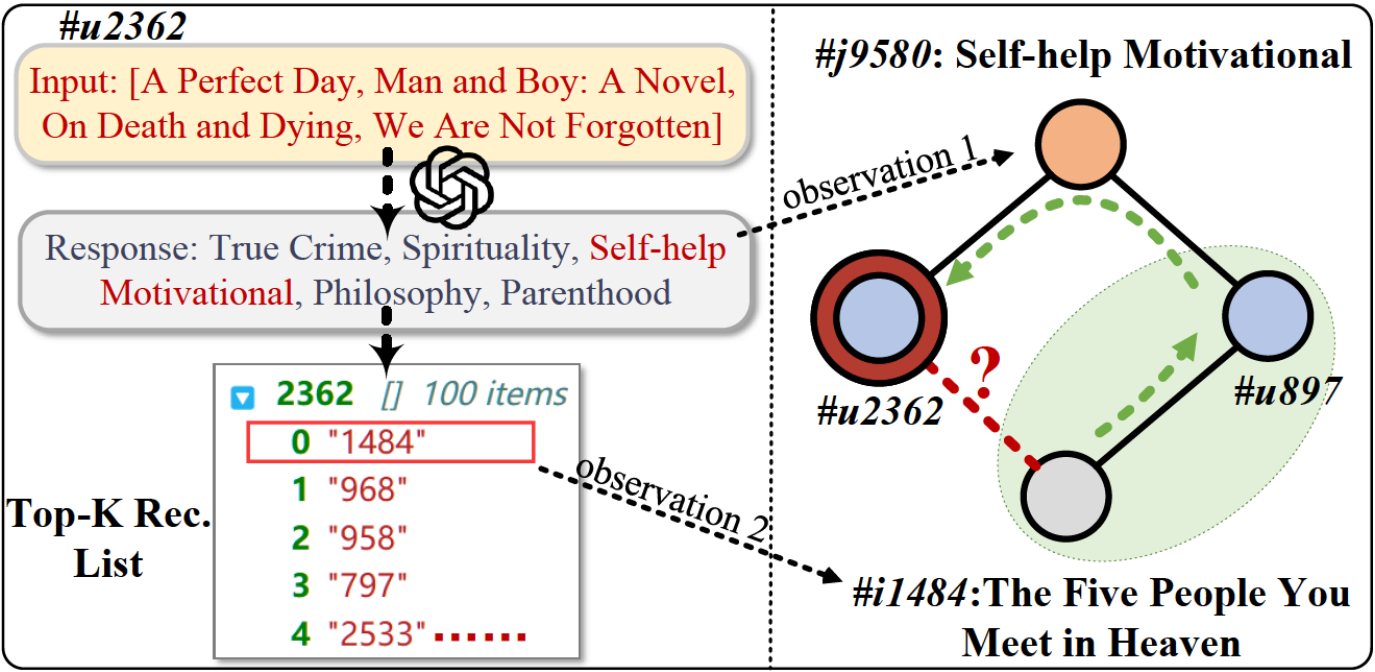}
	\caption{A real case in Book-Crossing dataset.}
	\label{fig:case_study}
 \end{center}
\end{figure}

\subsection{Case Study}

To intuitively assess the effectiveness of the user-side knowledge and the CIKG-based recommendation model, we select user $\#u2362$ from the Book-Crossing dataset, which exhibits the sparsest user-item interactions and item-side knowledge among the three datasets. As depicted in Fig.~\ref{fig:case_study}, the left panel illustrates the LLM-based interest knowledge generation process for user $\#u2362$ along with the top-$K$ recommendation list, where higher-ranked items indicate a greater likelihood of user interaction. The right panel demonstrates the process of capturing user similarity based on the shared interest $\#j9580$ between user $\#u2362$ and user $\#u897$ using the CIKG-based recommendation model. Item $\#i1484$, a book with a positive self-improvement narrative, appears in the test set for user $\#u2362$ and in the training set for user $\#u897$.

Based on Fig.~\ref{fig:case_study}, we draw the following two observations: \textbf{(1)} Our LLM-based user-side knowledge method effectively summarizes the interests of user $\#u2362$, as evidenced by the alignment between the potential interact item $\#i1484$ and interest $\#j9580$ (self-help motivational). This demonstrates that our proposed LLM-based user-side knowledge inference method effectively captures user interests. \textbf{(2)} Our CIKG-based model accurately predicts user $\#u2362$'s potential interaction with item $\#i1484$, as it is prominently listed in the top-K recommendation list for this user. This result suggests that our CIKG-based recommendation model effectively leverages user-side knowledge to identify higher-order user-item similarity relationships.

\begin{figure}[t]   
\begin{center}      
\includegraphics[width=0.45\textwidth]{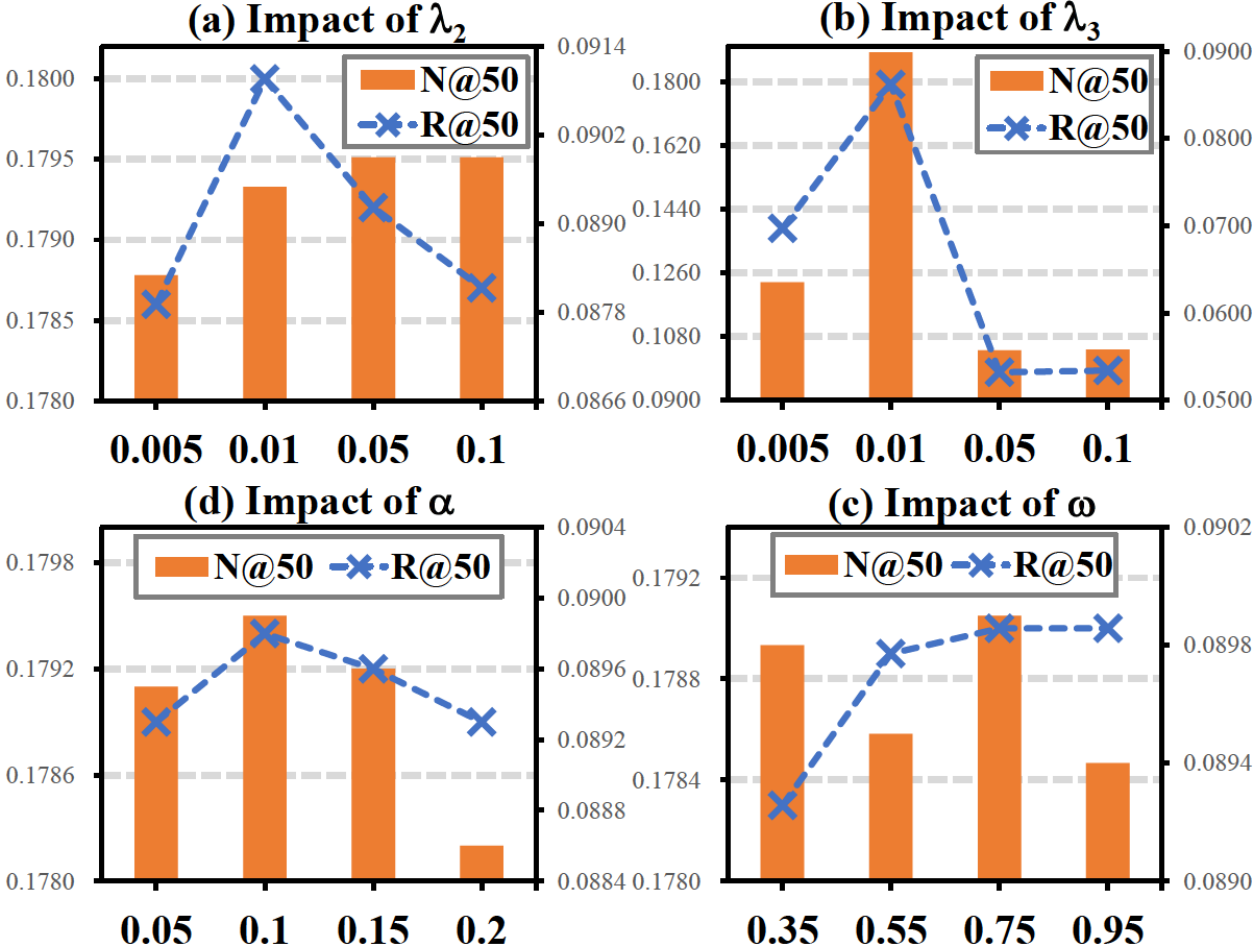}
	\caption{Hyperparameter sensitive experiments on Book-Crossing.}
	\label{fig:sensitive}
 \end{center}
\end{figure}

\subsection{Hyperparameter Sensitivity}
In this section, we investigate the influence of four critical hyperparameters: the initial mask rate $\alpha$, the maximum mask rate $\omega$, the user interest reconstruction loss weight $\lambda_2$, and the contrastive learning loss weight $\lambda_3$. The results of the hyperparameter sensitivity experiments on the Book-Crossing dataset are illustrated in Fig.\ref{fig:sensitive}. The primary vertical axis of the figures shows the $Recall@50$ results as a line chart, while the secondary vertical axis presents the $NDCG@50$ results as a bar chart. Based on Fig.\ref{fig:sensitive}, we derive the following observations: \textbf{(1)} Examining $\lambda_2$ and $\lambda_3$, which control the loss weights, reveals that model performance initially increases and then decreases as these parameters increase. Notably, a very high $\lambda_3$ leads to a significant decline in performance. This indicates that the selection of these hyperparameters should avoid extremes; they must be scaled appropriately during joint learning to achieve optimal performance. \textbf{(2)} Considering the initial mask rate $\alpha$ and the maximum mask rate $\omega$, we find that a small initial mask rate, around $0.1$, is beneficial for model training. The maximum mask rate $\omega$ can be quite large, even up to $0.95$. We believe that these hyperparameters should be adjusted according to the specific dataset and the particular LLM employed in our model.

\section{Conclusion}

In this paper, we address the issue of missing user-side structured knowledge, a problem often overlooked by knowledge-based recommender systems. We introduce a LLM-based method for generating user-side knowledge. Additionally, we propose a recommendation framework that effectively leverages the LLM-generated user-side knowledge. This framework includes a user interest reconstruction module and a contrastive learning module, which are designed to mitigate the challenges of noise and knowledge transfer, respectively.

\section{Acknowledgments}
This work was supported by the National Natural Science Foundation of China (Grant No. U24A20250 and 62176048, T2293771, 61673086, 42361144718), the Ministry of Education of Humanities and Social Science Project (No. 21JZD055), Sichuan Science and Technology Program (Grant No. 2023NSFSC1919), and the Sichuan Provincial Natural Science Foundation(Grant No. 2024NSFTD0042, 2024YFG0006, and 2024NSFSC0506). The funders had no role in the study design, data collection, analysis, decision to publish, or preparation of the manuscript.

\bibliography{aaai25}

\appendix
\section{Technical Appendix}

The appendix contains the following sections: the proof of Equation (4) from the main paper, the auxiliary information incorporation experiment, supplementary results on the sensitivity of additional hyperparameters, and descriptions of the baseline methods.

\section{Proof of Equation (4)}
Let $ h(q)\! =\! \delta_{lin}(q)\! -\! \delta_{exp}(q) $, thus $ h(0)\! =\! 0 $ and $ h(\Lambda)\! =\! 0 $. By taking the first and second derivatives of $ h(q) $, we get: 

\begin{align*}
\frac{d}{dq} h(q) &= \frac{d}{dq} (\delta_{lin}(q) - \delta_{exp}(q)) \\
        &= {\frac{d}{dq} \delta_{lin}(q)} - {\frac{d}{dq} \delta_{exp}(q)} \\
        &= \frac{(\omega-\alpha)}{\Lambda} - \frac{\alpha}{\Lambda}(\frac{\omega}{\alpha})^{\frac{q}{\Lambda}}{\ln(\frac{\omega}{\alpha})} \\
\frac{d^2}{dq^2} h(q) &= - \frac{\alpha}{\Lambda^2}(\frac{\omega}{\alpha})^{\frac{q}{\Lambda}}(\ln(\frac{\omega}{\alpha}))^2
\end{align*}

In our setup, \(\alpha\) and \(\omega\) represent the initial and maximum mask rates, both of which are greater than 0. Therefore, $ \frac{d^2}{dq^2} h(q) < 0 \text{ in the interval } [0, \Lambda], $ which means $ \frac{d}{dq} h(q) $ is monotonically decreasing in the interval \([0, \Lambda]\). Substituting \( q = 0 \) into \( \frac{d}{dq} h(q) \) and simplifying, we get:

\begin{align*}
\frac{d}{dq} h(q) \bigg|_{q = 0} &= \frac{(\omega - \alpha)}{\Lambda} - \frac{\alpha}{\Lambda}(\frac{\omega}{\alpha})^{\frac{0}{\Lambda}}\ln(\frac{\omega}{\alpha}) \\
&= \frac{\alpha}{\Lambda}((\frac{\omega}{\alpha} - 1) - \ln(\frac{\omega}{\alpha})) 
\end{align*}

Since \(\omega > \alpha\), it follows that \(\frac{\omega}{\alpha} > 1\). It is easy to prove that $ (\frac{\omega}{\alpha} - 1) - \ln(\frac{\omega}{\alpha}) $ is always greater than 0 when \(\frac{\omega}{\alpha} > 1\). Consequently, $ \frac{d}{d q} h(q) \bigg|_{q = 0} > 0 $. Then substituting \( q = \Lambda \) into \( \frac{d}{dq} h(q) \) and simplifying, we get:

\begin{align*}
\frac{d}{dq} h(q) \bigg|_{q = \Lambda} &= \frac{(\omega - \alpha)}{\Lambda} - \frac{\alpha}{\Lambda} \left( \frac{\omega}{\alpha} \right)^{\frac{\Lambda}{\Lambda}} \ln \left( \frac{\omega}{\alpha} \right) \\
&= \frac{\alpha}{\Lambda}(\frac{\omega}{\alpha}(1 - \ln(\frac{\omega}{\alpha})) - 1)
\end{align*}

Similarly, it is easy to prove that $ \frac{\omega}{\alpha} \left( 1 - \ln \left( \frac{\omega}{\alpha} \right) \right) - 1 $ is always less than 0 when \(\frac{\omega}{\alpha} > 1\). Consequently, $ \frac{d}{d q} h(q) \bigg|_{q = \Lambda} < 0 $. Thus, we can conclude the following:

\[
\exists q_0     \text{ s.t. } 
\begin{cases} 
\frac{d}{dq} h(q) > 0 & \text{for } q \in [0, q_0) \\
\frac{d}{dq} h(q) < 0 & \text{for } q \in (q_0, \Lambda]
\end{cases}
\]

\begin{figure}[t]   
\begin{center}      
\includegraphics[width=0.35\textwidth]{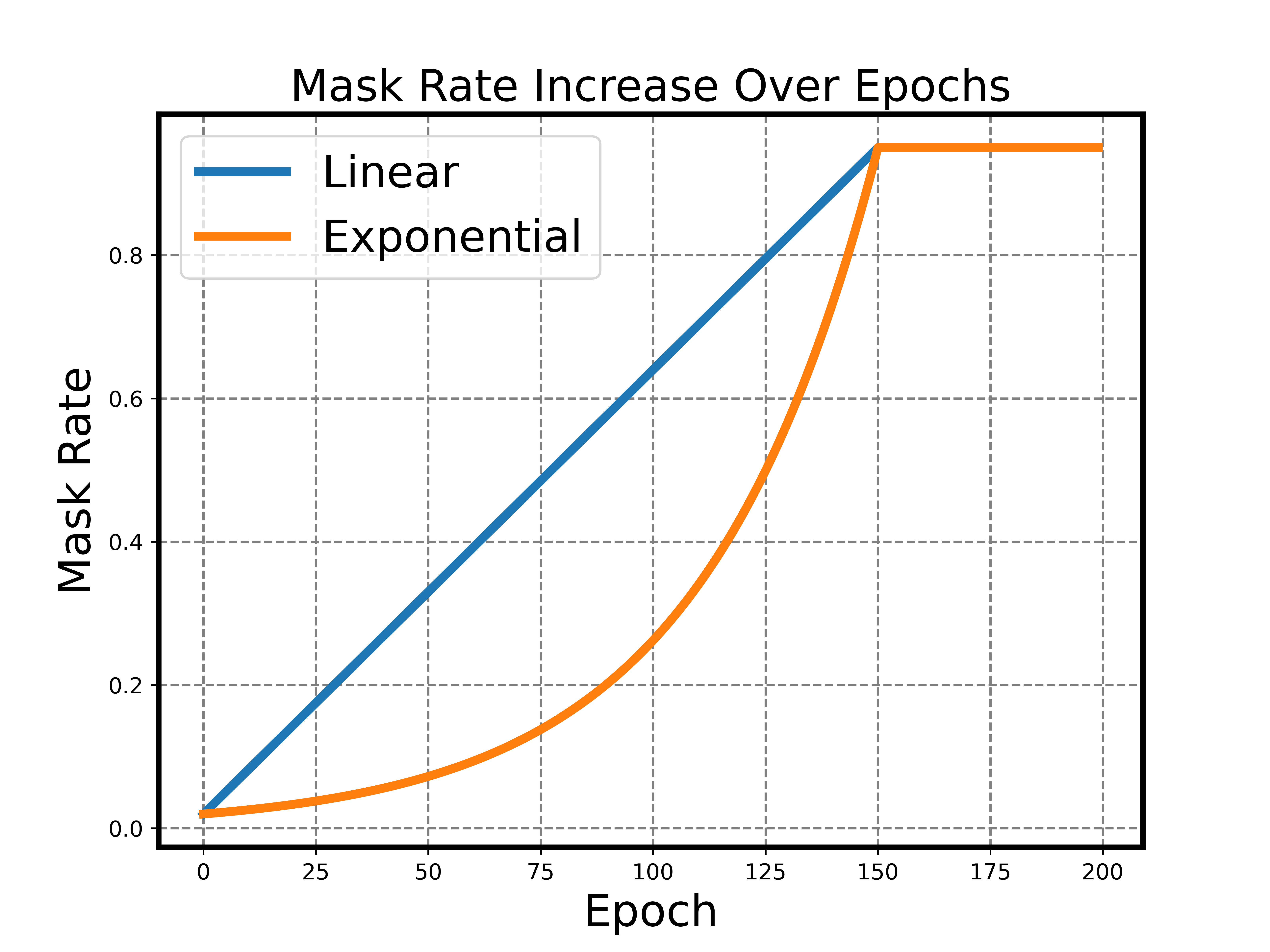}
	\caption{A mask rate scheduling line chart example, where $\alpha=0.02,\ \omega=0.95, \Lambda=150$.}
	\label{fig:chart}
 \end{center}
\end{figure}

Therefore, $ h(q) $ is monotonically increasing for $ q \in [0, q_0) $, $ h(q) $ is monotonically decreasing for $ q \in (q_0, \Lambda] $. Since \( h(0)\! =\! 0 \) and \( h(\Lambda)\! =\! 0 \), \( h(q) \) is always greater than 0 in the interval \((0, \Lambda)\). Thus, equation (4) holds, and the proof is complete. A mask rate scheduling line chart example is provided in Fig.~\ref{fig:chart}.

\begin{table*}[t]
  \centering
  \tabcolsep=0.5mm
  \renewcommand\arraystretch{1.5}
  \caption{LightGCN benchmark with variant auxiliary information.}
    \begin{tabular}{lrrrrrrrrrrrr}
    \toprule
          & \multicolumn{4}{c}{dbbook2014} & \multicolumn{4}{c}{book-crossing} & \multicolumn{4}{c}{movielens-1m} \\
    \midrule
          & \multicolumn{1}{l}{R@50} & \multicolumn{1}{l}{N@50} & \multicolumn{1}{l}{R@100} & \multicolumn{1}{l}{N@100} & \multicolumn{1}{l}{R@50} & \multicolumn{1}{l}{N@50} & \multicolumn{1}{l}{R@100} & \multicolumn{1}{l}{N@100} & \multicolumn{1}{l}{R@50} & \multicolumn{1}{l}{N@50} & \multicolumn{1}{l}{R@100} & \multicolumn{1}{l}{N@100} \\
    \cmidrule(r){2-5} \cmidrule(r){6-9} \cmidrule(r){10-13}
    LightGCN$_{CG}$ & 0.4214 & 0.223 & 0.5358 & 0.2481 & \underline{0.1607} & 0.0765 & \underline{0.2246} & 0.0911 & 0.4086 & 0.3977 & 
    \underline{0.5574} & 0.4432 \\
    \hline
    LightGCN$_{CIG}$ & 0.4240 & \textbf{0.2283} & \underline{0.5395} & \textbf{0.2537} & \textbf{0.1611} & \underline{0.0773} & \textbf{0.2247} & \underline{0.0916} & 0.4079 & 0.3982 & 0.5571 & 0.4439 \\
    LightGCN$_{CKG}$ & \textbf{0.4273} & 0.2276 & 0.5394 & 0.2524 & 0.1602 & \textbf{0.0790} & 0.2197 & \textbf{0.0928} & \textbf{0.4097} & \underline{0.3985} & \textbf{0.5584} & \underline{0.4441} \\
    LightGCN$_{CIKG}$ & \underline{0.4272} & \underline{0.2279} & \textbf{0.5428} & \underline{0.2534} & 0.1603 & 0.0759 & 0.2231 & 0.0901 & \underline{0.4096} & \textbf{0.3995} & \underline{0.5574} & \textbf{0.4446} \\
    \bottomrule
    \end{tabular}%
  \label{tab:lgn_exp}%
\end{table*}%

\section{Auxiliary Information Incorporation Experiment}

To evaluate the effectiveness of incorporating item-side information (KG) and our proposed user-side information (IG) into the recommendation domain (CG), we employ LightGCN \cite{DBLP:conf/sigir/0001DWLZ020} as the baseline model. We fix all hyperparameters, including random seeds, and use graphs containing various types of auxiliary information to assess the impact of this additional information on model performance. The experimental results are presented in Tab.~\ref{tab:lgn_exp}. In this context, LightGCN$_{CG}$ refers to the original model using only collaborative data, LightGCN$_{CIG}$ denotes the variant augmented with LLM-generated user-side information, LightGCN$_{CKG}$ incorporates item-side information, and LightGCN$_{CIKG}$ integrates both user-side and item-side information. The key observations are as follows:

\begin{itemize}
\item Models incorporating auxiliary information consistently outperform the baseline model (LightGCN$_{CG}$) across all three datasets. This aligns with our hypothesis that the inclusion of additional information enables GNN-based models to capture higher-order collaborative signals, thereby enhancing performance.
\item Among the variants incorporating auxiliary information, both user-side and item-side additions improve model performance, although no single variant consistently outperforms the others across all metrics and datasets. This result supports the effectiveness of the user-side information generated by our LLM. However, the lack of a dominant variant suggests that noise in the auxiliary information, combined with the absence of an effective knowledge transfer module in the original LightGCN, limits the model's ability to fully leverage this information. Thus, our findings highlight the necessity of our proposed anti-noise and knowledge transfer modules.
\end{itemize}

\begin{figure}[t]   
\begin{center}  
\subfigure{
\includegraphics[width=0.5\textwidth]{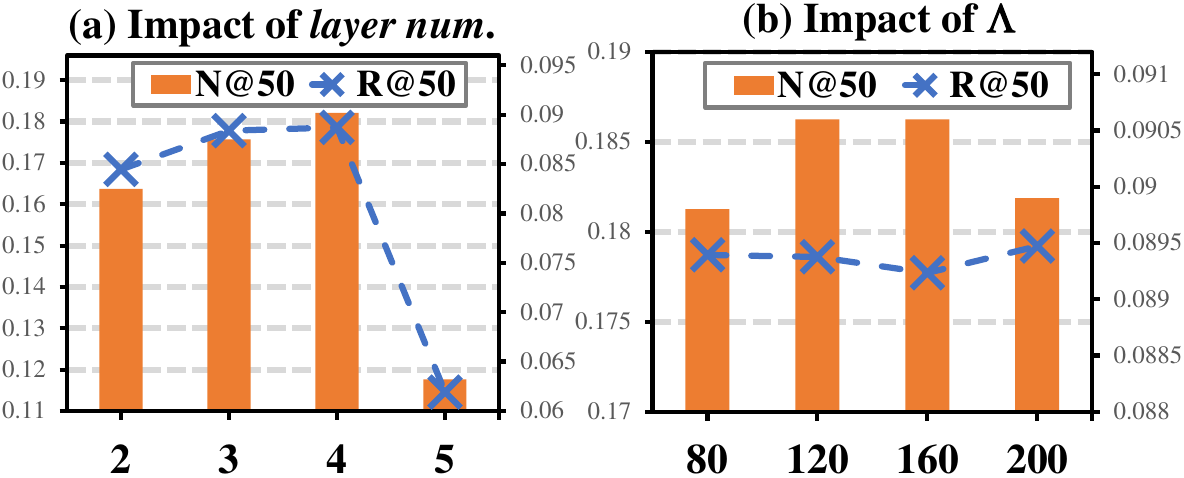}
    }
\subfigure{
\includegraphics[width=0.3\textwidth]{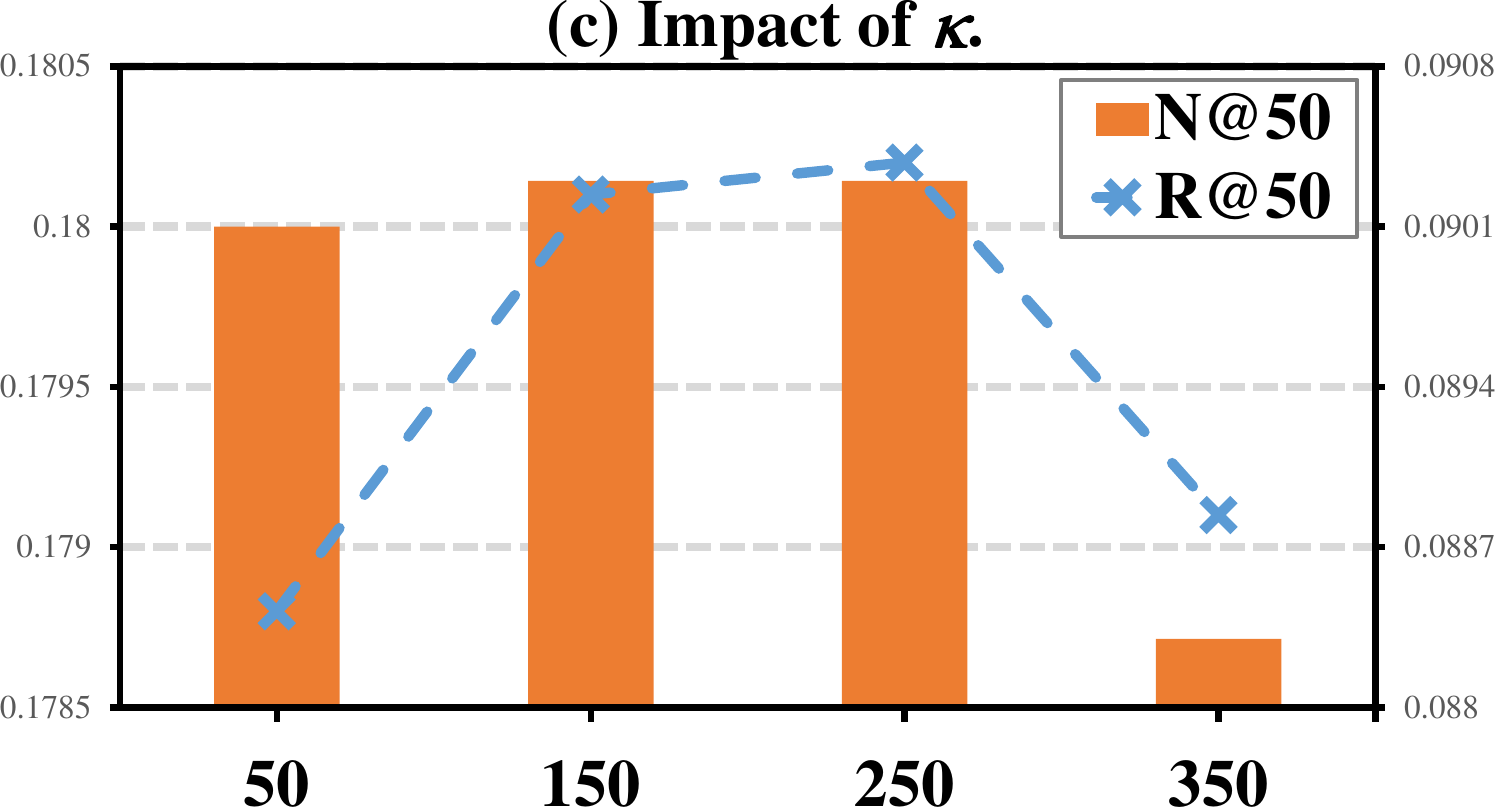}
    }
    \caption{Supplementary Hyperparameter sensitive experiments on book-crossing.}
\label{fig:sensitive}
 \end{center}
\end{figure}

\section{Supplementary Hyperparameter Sensitive Experiments}
To further investigate the impact of different hyperparameter settings on our model's performance, we conduct sensitivity experiments focusing on the number of graph encoder layers, the cluster number of user interests ($\kappa$) and the number of epochs ($\Lambda$) required to reach the maximum mask probability. The results are presented in Fig.~\ref{fig:sensitive}. From these experiments, we draw two key observations: 
\begin{itemize}
    \item As shown in Fig.~\ref{fig:sensitive}(a), the model performs optimally with $3$ and $4$ layers. However, increasing the number of layers beyond this range results in a significant drop in performance, likely due to over-smoothing.
    \item Regarding $\Lambda$, our findings indicate that the number of epochs needed to reach the maximum mask probability should be carefully balanced; values that are either too high or too low can adversely affect performance.
    \item The number of clusters $\kappa$ significantly influences model performance. An insufficient number of clusters can result in over-smoothing by GNNs, whereas an excessive number may overly fragment user interests. In practice, this hyperparameter should be determined based on the characteristics of the specific dataset.
\end{itemize}

\section{Details of Baseline Methods}
In this section, we provide the detailed introduction of the baselines methods utilized in this paper.
\begin{itemize}
    \item LightGCN \cite{DBLP:conf/sigir/0001DWLZ020} is a simplified Graph Convolutional Network model for collaborative filtering that focuses solely on neighborhood aggregation. It linearly propagates user and item embeddings across the user-item interaction graph and uses a weighted sum of the embeddings from all layers as the final embedding. This streamlined approach excludes feature transformation and nonlinear activation, making the model easier to implement and train.
    \item SGL \cite{DBLP:conf/sigir/WuWF0CLX21} enhances GCN-based recommendation models by introducing an auxiliary self-supervised task that improves the accuracy and robustness of node representations. It generates multiple views of each node through node dropout, edge dropout, and random walks, and maximizes agreement between different views of the same node. This approach is implemented on the LightGCN model, helping to mitigate issues related to high-degree nodes and noisy interactions in recommendation tasks.
    \item SimGCL \cite{DBLP:conf/sigir/YuY00CN22} discards the use of graph augmentations and instead generates contrastive views by adding uniform noise to the embedding space. This method works by learning more evenly distributed user/item representations, which helps to mitigate popularity bias. The approach simplifies the traditional CL pipeline while improving recommendation accuracy and training efficiency.
    \item CKE \cite{DBLP:conf/kdd/ZhangYLXM16} is a framework that integrates collaborative filtering with semantic representations derived from a knowledge base. It extracts structural representations using the TransR heterogeneous network embedding method, textual representations with stacked denoising auto-encoders, and visual representations using stacked convolutional auto-encoders. These semantic representations are jointly learned with collaborative filtering to enhance recommendation quality.
    \item CFKG \cite{DBLP:journals/algorithms/AiACZ18} integrates knowledge-base embedding with collaborative filtering to improve recommendation accuracy and generate personalized explanations. It learns representations for users and items by embedding heterogeneous entities from a structured knowledge base, preserving their relationships with external knowledge. A soft matching algorithm is then used to generate personalized explanations for recommended items based on the embedded knowledge base.
    \item KGAT \cite{DBLP:conf/kdd/Wang00LC19} integrates a knowledge graph with user-item interactions to improve recommendation accuracy by modeling high-order relations between items and their attributes. It propagates embeddings through a hybrid structure of the knowledge graph and user-item graph, refining node embeddings using recursive neighbor propagation. An attention mechanism is applied to weigh the importance of these neighbors, enhancing the interpretability and effectiveness of the model.
    \item KGIN \cite{DBLP:conf/www/WangHWYL0C21} enhances recommender systems by modeling user-item interactions at a fine-grained level of intents, using auxiliary item knowledge. It represents each intent as a combination of knowledge graph relations and employs a new GNN-based information aggregation scheme to recursively integrate relational paths, preserving long-range connectivity semantics. This approach enables the extraction and encoding of user intents into the representations of users and items, improving model capability and interpretability.
    \item KGCL \cite{DBLP:conf/sigir/YangHXL22} addresses challenges in knowledge graph-enhanced recommendation systems caused by sparse supervision and noisy data. KGCL employs a knowledge graph augmentation schema to reduce noise and improve item representations, and integrates additional supervision signals into a cross-view contrastive learning paradigm. This approach enhances the accuracy of user preference representation by focusing on unbiased user-item interactions.
    \item MCCLK \cite{DBLP:conf/sigir/Zou0MWQ0C22} introduces a multi-level cross-view contrastive learning mechanism for knowledge graph-aware recommendation, addressing the sparse supervised signal problem in GNN-based models. It considers three distinct graph views: a collaborative view (user-item graph), a semantic view (item-entity graph), and a structural view (user-item-entity graph), and performs contrastive learning across these views at both local and global levels to mine comprehensive graph features. Additionally, MCCLK employs a k-Nearest-Neighbor (kNN) module in the semantic view to capture crucial item-item semantic relations often overlooked in prior approaches.
    \item KGRec \cite{DBLP:conf/kdd/YangHXH23} is a self-supervised rationalization method for knowledge-aware recommender systems that uses an attentive knowledge rationalization mechanism to generate scores for knowledge triplets. It integrates generative and contrastive self-supervised tasks through rational masking, where high-scoring knowledge is masked and then reconstructed to highlight key connections. Additionally, KGRec employs a contrastive learning task to align knowledge and user-item interaction signals while masking potential noisy edges based on the rational scores.
\end{itemize}

\end{document}